\documentclass{webofc}
\usepackage[varg]{txfonts}   
\begin{document}

\title{Exact strangeness conservation in heavy ion collisions\footnote{In memory of  Jean Cleymans}}
%
%

\author{\firstname{Krzysztof} \lastname{Redlich}\inst{1}\fnsep\thanks{\email{krzysztof.redlich@uwr.edu.pl
    }} \and
        \firstname{Natasha} \lastname{Sharma}\inst{2}\fnsep\thanks{\email{natasha.sharma@cern.ch}} 
}

\institute{Institute of Theoretical Physics, University of Wroc\l aw, 50-204  Wroc\l aw, Poland
\and
          Indian Institute of Science Education and Research (IISER) Berhampur, 760010 Odisha, India
          }

\abstract{%
We investigate the increase in strangeness production  with charged particle
multiplicity ($dN_{ch}/dy$) seen by the ALICE collaboration at CERN 
in p-p, p-Pb and Pb-Pb collisions using the  hadron resonance gas model.
The strangeness canonical ensemble is used taking into account
the interactions among hadrons using S-matrix
corrections  based on known phase shift analyses.
We  show the essential role of constraints due to the  exact conservation of strangeness which is instrumental to describing observed  features of strange particle yields  and their scaling with $dN_{ch}/dy$. Furthermore, 
the results on comparing the hadron resonance gas model with and without
S-matrix corrections, are presented.
We observe that the interactions introduced by the phase shift
analysis via the S-matrix formalism are essential for a better  
description of the yields data independent of collision system. This work is based on our long-lasting collaboration  and most recent publications with Jean Cleymans\footnote{https://home.cern/news/obituary/cern/jean-willy-andre-cleymans-1944-2021}. 
}
\maketitle
\section{Introduction}
\label{intro}
The experimental particle yield data from heavy-ion collisions across
different experiments and broad range of energies have been shown to
originate from a thermal production
\cite{Andronic:2017pug,Andronic:2014zha,Becattini:2010sk,Andronic:2018qqt,Chatterjee:2015fua,Das:2016muc,Abelev:2013haa}. These
data have been successfully explained by the hadron gas model (HRG)
with a common freezeout temperature $T_f$ and chemical potentials
$\vec\mu_f$  associated with the conserved charges. It has been
shown that the freezeout temperature obtained from the HRG
conincides with the chiral crossover temperature from lattice QCD
(LQCD) at vanishingly small finite $\mu_B$ \cite{Bazavov:2017dus}.


   

It is argued that the S-matrix scheme can improve the HRG model in
approximating the QCD partition function in the hadronic phase and hence
can provide a more accurate description of the measured particle
yields \cite{Dashen:1969ep,Venugopalan:1992hy,Weinhold:1997ig,Lo:2017lym,Dash:2018mep,Dash:2018can}.
In this work, we apply the S-matrix extended HRG model to  analyze data obtained by the  ALICE 
collaboration on charged-particle multiplicity in p-p collisions at 7 TeV~\cite{ALICE:2017jyt} as well as
13 TeV \cite{Acharya:2019kyh},  
in  p-Pb collisions at 5.02 TeV~\cite{Abelev:2013haa,Adam:2015vsf} and in  Pb-Pb collisions
at 2.76 TeV~\cite{Abelev:2013vea,Abelev:2013xaa,Abelev:2013zaa},  all in  the central region of rapidity.  

We estimate the yields of (multi-)strange hadrons and discuss strangeness suppression as a function of $dN_{ch}/d\eta$ and strangeness content of hadrons. We formulate the HRG model in the canonical ensemble of strangeness conservation and account for differences between the fireball volume  at midrapidity $V_A$  and the correlation volume $V_C$ required for exact global strangeness conservation
\cite{Hamieh:2000tk,Hagedorn:1984uy,Satz:2017ltg,Castorina:2013mba}.
We  include the S-matrix corrections to proton production by using the  empirical phase shifts of $\pi N$ scattering 
and the contribution of hyperons to proton yield
and corrections to the hyperon yields,  employing an existing coupled-channel study involving  $\pi \Lambda$,  and $\pi \Sigma$  interactions 
in the $S = -1$ sector \cite{Cleymans:2020fsc}.

\section{Strangeness production and HRG model}
\label{sec-1}
In the framework of HRG model with the grand canonical ensemble,
the quantum numbers are conserved on  average and  are  implemented
using the corresponding  chemical potentials $\vec\mu$ linked to
conserved charges $\vec Q$. At the LHC energies, all chemical
potentials $\vec\mu$ are set to zero, i.e. the  system is charge
neutral.  Such a  model has been  applied to quantify thermalization
and particle production  in most central heavy-ion collisions at the
LHC. 
At low multiplicity events or low collisions energies, a thermal description requires  exact implementation of charge  conservation which is usually  described in  the C-ensemble~\cite{Hagedorn:1971mc,Hagedorn:1984uy,Cleymans:1990mn,Cleymans:1998yb,Hamieh:2000tk,Ko:2000vp,BraunMunzinger:2003zd}.

We focus on the canonical ensemble with exact strangeness conservation (total strangeness $S=0$) and assume all other quantum numbers are conserved on average in the GC ensemble with $\vec \mu=0$. 
The strangeness canonical partition function can  be expressed as  a series of  Bessel functions \cite{BraunMunzinger:2001as,BraunMunzinger:2003zd,Cleymans:1990mn,Hamieh:2000tk}, 
\begin{eqnarray}
Z^C_{S=0}&=& \sum_{n,p=-\infty}^{\infty}
 I_n(S_2) I_p(S_3) I_{-2n-3p}(S_1). 
\label{eq7}
\end{eqnarray}

The  mean multiplicity in the canonical ensemble of particle $k$ carrying strangeness $s$ in  a given experimental acceptance $A$,  are obtained as
\begin{eqnarray}
\label{equ9}
\langle N_k^s\rangle_A =& V_A \, n_k^s(T) \,  \frac{1}{Z_{S=0}^C}  \times \sum_{n,p=-\infty}^{\infty} \, I_n(S_2) I_p(S_3) I_{-2n-3p- s}(S_1).
\end{eqnarray}

The particle yields are obtained 
using the temperature ($T$) and  two volume parameters: the volume  of
the system in the acceptance  $V_A$ and the correlation volume $V_C$
of  global strangeness  conservation which appear in the arguments of
the Bessel functions.
To get the total multiplicity of a given strange particle in the C
ensemble resonance decays have to be added.
To identify the contribution and dependence of various terms in
Eq.~\eqref{equ9} on strange particle quantum number $s$, only leading
terms can be considered ($p=n=0$), so 
\begin{align}
\label{equ10}
\langle N_k^s\rangle_A \simeq V_A \, n_k^s(T) \, \frac{I_{s}(S_1)}{I_{0}(S_1)}.
\end{align}
The ratio of ${I_{s}(S_1)/ I_{0}(S_1)}$ is the suppression factor
which  decreases   with increasing  $s$  and with decreasing thermal
phase-space occupied by strange particles.
These correspond to the strangeness canonical suppression that have been introduced  \cite{Hamieh:2000tk} to describe  thermal production of multi(strange) hadrons   in heavy-ion collisions. 
In this work, we apply the HRG model with the C-ensemble to quantify
the production of (multi-)strange hadrons and their 
behavior with charged particle multiplicity. We modify the HRG model
with a more complete implementation of interactions within the
S-matrix formalism. 

\section{S-matrix formalism and HRG}
The various interactions among hadrons modify the density of states (DOS) of a thermal system and hence the thermal abundances of hadron states~\cite{Dashen:1969ep,Venugopalan:1992hy,Lo:2017sde,Lo:2020phg}. 
In the scattering matrix (S-matrix) formalism, an effective spectral
function $B(M)$ ($M$ represents center-of-mass energy of the system)  describing the DOS can be computed from the
S-matrix~\cite{Dashen:1969ep}. It summarizes the various interactions
among the scattering channels. 
As the energy increases, new interaction channels open and scattering
becomes inelastic. 
The channel-specific spectral
function $B_a(M)$ describes the energy dependent component of the full
$B(M)$ for a given channel $a$.
The channel yield, e.g., from a resonance decaying into multiple final states,
can be obtained as
\begin{align}
n_a(T) =  \int_{m_{th}}^\infty \frac{dM}{2\pi} B_a(M) \, n^{(0)}(T,M),
\end{align}
where $n^{(0)}$ is the ideal gas formula for the particle density. 
\begin{figure}[ht]
\sidecaption
\centering
\includegraphics[width=0.65\linewidth]{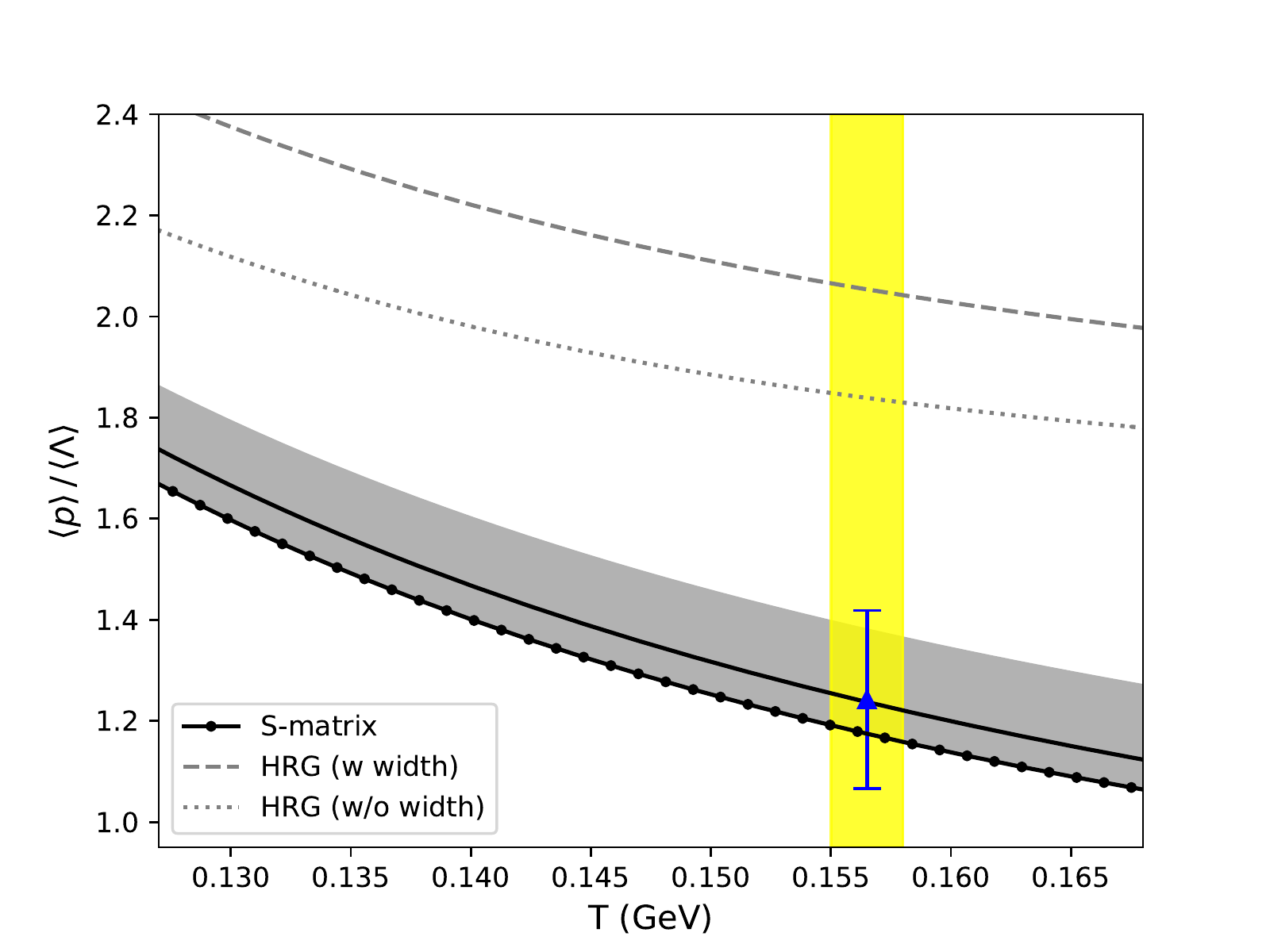}
	\caption{ Proton to  $\Lambda+\Sigma^0$-baryons yield ratio is shown as a function of temperature. 
	The predictions from the S-matrix scheme  and from the HRG scheme (with and without width) are presented as line, dashed line and dotted line respectively. The blue solid triangle represents the experimental value measured by the ALICE collaboration for the most central Pb-Pb collisions (at $2.76$ TeV).
        }
\label{fig:fig4}
\end{figure}

 We focus on computing the hadron yields of protons and
$\Lambda+\Sigma^0$-baryons. 
Following previous studies for protons, 
we employ the empirical phase shifts from the GWU/SAID PWA on the $\pi N$ scattering~\cite{Workman:2012hx}. 
In addition to this, we implement a $\pi \pi N$ background contribution based on
the LQCD computation of the baryon charge-susceptibility $\chi_{BQ}$.
For the strange ($\vert S \vert =1$) baryon system, we 
employ an existing coupled-channel model involving $\bar{K} N$, $\pi \Lambda$, and $\pi \Sigma$ interactions. 
The S-matrix scheme for $\vert S \vert = 1$ hyperons is based on
Ref.~\cite{hyp_ccm} without any extra tuning.

It is observed that the HRG model overestimates the yields of protons
and  underestimates the yields of $\Lambda+\Sigma^0$-baryons. As a
result, the ratio of the yields of protons to
$\Lambda+\Sigma^0$-baryons is  larger than the measured
experimental ratio.  Figure~\ref{fig:fig4} shows the comparison of
ratio of protons to $\Lambda+\Sigma^0$ yields measured by the ALICE
collaboration for the most central Pb-Pb collisions (at $2.76$ TeV),
with the corresponding predictions from the HRG model and from the
S-matrix scheme.
It is observed that the result obtained from the measured hadron
yields favor the S-matrix prediction and hence illustrates the
robustness of this method. 

\section{Comparison of results with ALICE data}

We have used the latest version of THERMUS~\cite{Wheaton:2004qb} and
extended it to include the S-matrix approach.
As a first step, we made a fit for each multiplicity bin of pp collisions at $\sqrt{s}$ = 7 TeV~\cite{ALICE:2017jyt}
by keeping the number of parameters to a minimum.
Two temperature values were chosen: $T_f$ = 156.5 MeV and $T_f$ = 160 MeV
based on the results from LQCD \cite{Bazavov:2018mes} and the recent HRG model analysis of ALICE data for central Pb-Pb collisions \cite{Andronic:2018qqt}.
The strangeness suppression factor $\gamma_s$ is fixed to 1 ($\gamma_s
= 1$) as motivated by fits  in central Pb-Pb collisions,.
The chemical potentials due to  conservation of  baryon number and
electric charge are set to zero for LHC energy.
Thus, in the SCE  only two parameters remained, the volume of the system in the experimental acceptance $V_A$  and the canonical volume $V_C$ which quantifies  the range  of exact strangeness conservation.

\begin{figure*}[ht]
\centering
\includegraphics[width=0.71\linewidth]{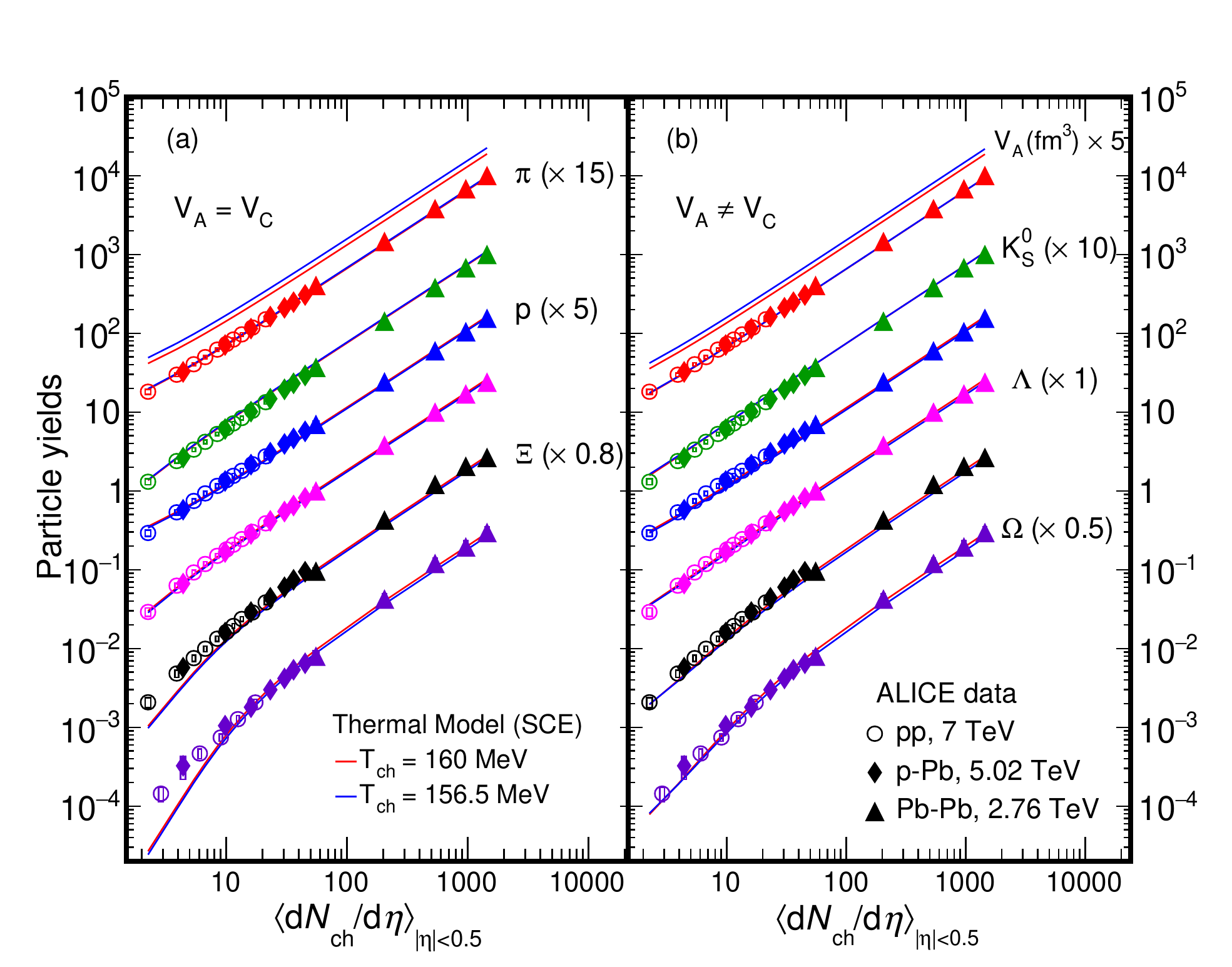}
	\caption{Left: Yields for $V_A=V_C$. Right: Yields for $V_A\neq V_C$,
The two	top lines are the fitted volumes $V_A$ (x5) in fm$^3$. The particle yields are indicated together with the multiplicative factor used to separate the yields. The solid blue lines have been calculated for $T$ = 156.5 MeV while the solid red lines have been calculated for $T$ = 160 MeV. 
The values of the volumes used have been parametrized empirically (see text for more details). 
}
\label{fig:fig5}
\end{figure*}

Figure \ref{fig:fig5} (left) shows the hadrons yields calculated using the SCE for different charged particle multiplicities $dN_{ch}/d\eta $  for a single volume  $V=V_A\simeq V_C$. 
For comparison, the experimental results from the ALICE collaboration are also shown as symbols.
 The SCE model 
qualitatively explains the hadron yields data even with a single
volume  parameter. For large $dN_{ch}/d\eta >  100$ all hadron yields,
as well as  extracted $V$,   depend   linearly  on  charged particle
rapidity density. However,  for lower $dN_{ch}/d\eta$ this dependence
is clearly non-linear for strange particles  due to strangeness
canonical  suppression which increases with the strangeness content of
particles. At the quantitative level, however, as seen  in
Fig. \ref{fig:fig5} (left), using a single volume leads to too much suppression at small charged particle multiplicities,  particularly for $S=-2$ and $S=-3$ baryons.
This result is consistent with the  previous observation, that a single volume canonical model implies   too strong  strangeness suppression in low multiplicity events~\cite{Adam:2015vsf}. 

We have also performed the SCE model fit  to data  with two independent volume parameters as shown in Fig. \ref{fig:fig5} (right). In general, strangeness conservation relates to the full phase-space whereas particle yields are measured in some acceptance window. Thus, the strangeness canonical volume  parameter  $V_C$ can be larger than the fireball volume $V_A$,   restricted to a given acceptance. The resulting yields  exhibit  much better agreement with data by decreasing strangeness suppression at lower multiplicities due to  larger value of $V_C$ than $V_A$.  



\begin{figure}[h]
\centering
\includegraphics[width=0.51\linewidth]{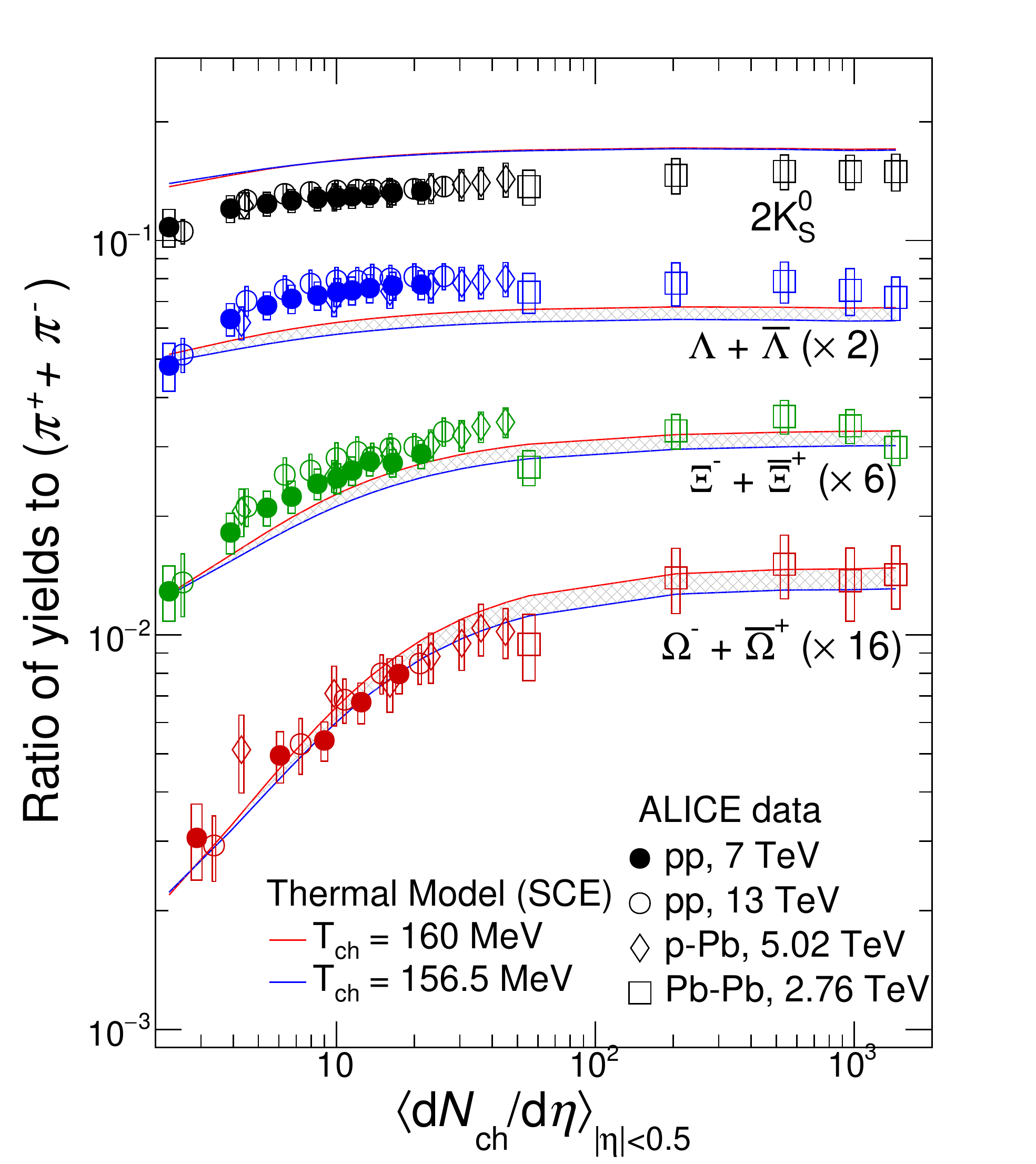}
\caption{Strange and multi-strange particles to pions yield ratios versus charged particle multiplicity. The symbols show experimental data and the lines represent the SCE model results obtained using $V_A\neq V_C$.}
\label{fig:fig8}       
\end{figure}

The pion yields data  are slightly underestimated by the model
points while the kaons are overestimated. This has implications for
the kaon to pion ratio discussed further below.
The strangeness suppression effect and its CSE model description are
important to describe the dependence 
of  particle yields  on the fireball   volume $V_A$.
The ratio of strange particle and  pion yields,   is shown in Fig.~\ref{fig:fig8}. 
This ratio has been discussed prominently by the ALICE collaboration~\cite{ALICE:2017jyt} where 
a comparison with other model calculations was presented. The SCE  model introduced  here compares very favorably to the ones discussed in~\cite{ALICE:2017jyt}.
The underestimation of the pion yield is responsible for the larger discrepancy in the kaon to pion ratio.

\begin{figure*}[ht]
\sidecaption
\includegraphics[width=0.51\linewidth]{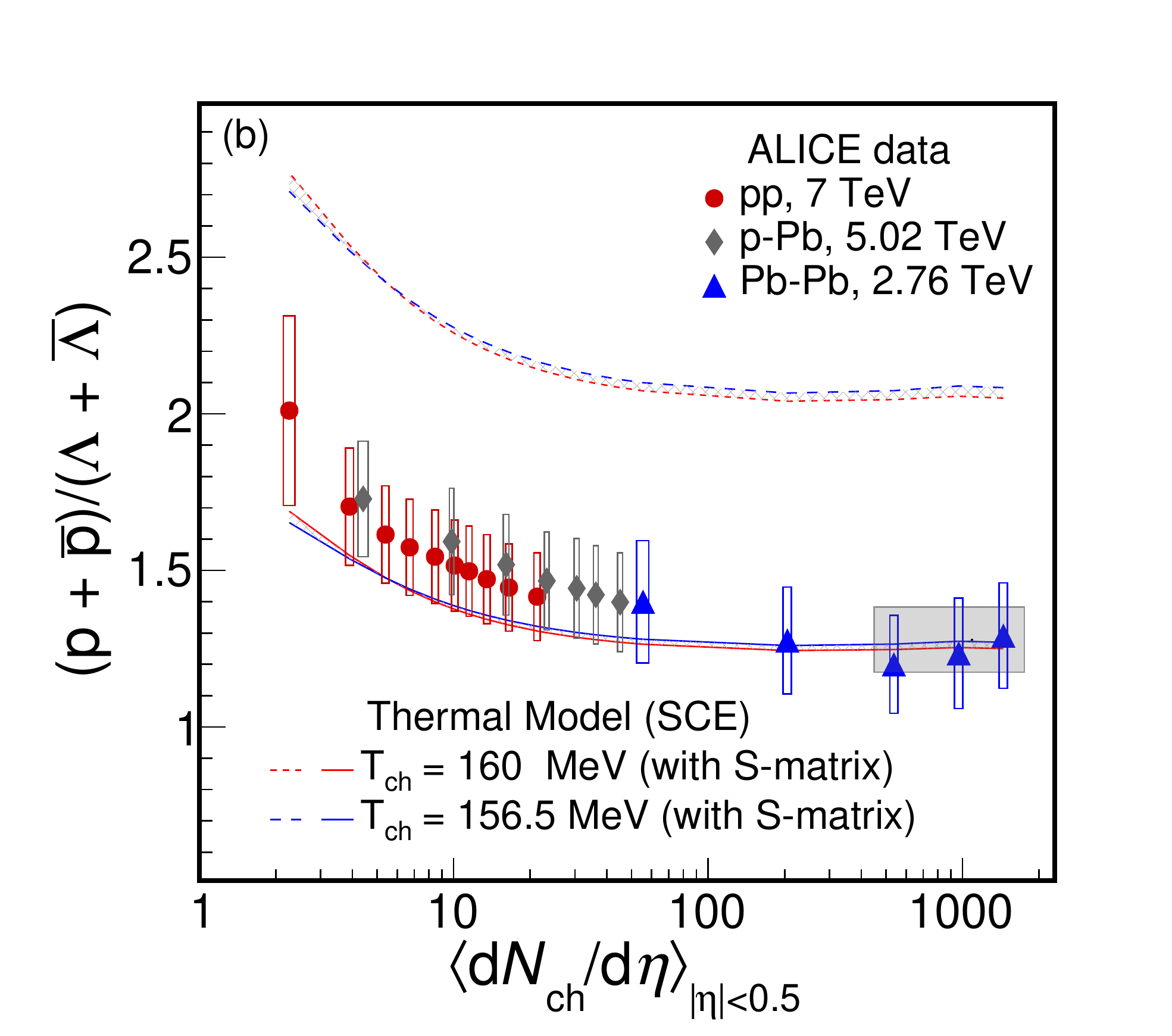}
\caption{Proton to Lambda ratio as a function of charged particle multiplicity is shown. 
The lines represent predictions from the HRG model including the S-matrix corrections  while the dotted lines represent predictions from the HRG model without S-matrix corrections at temperature T = 160 MeV and 156.5 MeV.
The solid symbols represent the experimental data measured by the ALICE collaboration\cite{ALICE:2017jyt,Acharya:2019kyh,Abelev:2013haa,Adam:2015vsf,Abelev:2013vea,Abelev:2013xaa,Abelev:2013zaa}.
}
\label{fig:fig9}
\end{figure*}

The importance of the S-matrix description of proton and hyperon
yields can be seen with data from the central Pb-Pb collisions 
by comparing measured $p/\Lambda$ ratio with the S-matrix results as shown in Fig.~\ref{fig:fig9}. 
The lines represent predictions from the HRG model including the S-matrix corrections  while the dotted lines represent predictions from the HRG model without S-matrix corrections at temperature T = 160 MeV and 156.5 MeV.
The shaded region (in gray) shows the theoretical predictions spanned by different stages of the S-matrix improvement: from including only elastic scatterings of ground state hadrons (upper limit) to the full list of interactions (lower limit). The solid symbols represent the experimental data measured by the ALICE collaboration\cite{ALICE:2017jyt,Acharya:2019kyh,Abelev:2013haa,Adam:2015vsf,Abelev:2013vea,Abelev:2013xaa,Abelev:2013zaa}.
The ratio is very sensitive to the S-matrix corrections which amount
to a reduction of  proton yield by a factor of 0.75 and a 1.24
enhancement of the $\Lambda$ yield.
The thermal model without S-matrix corrections exhibits  large deviations from the  $p/\Lambda$  ratio data. 
 We also found that the ratios of strange and multi-strange particles from SCE model describe the ALICE data very accurately.  





\section{Summary}

We  have studied   the  effect  of  global  strangeness quantum number
conservation  on  strangeness production  in heavy-ion  and elementary
collisions in a given acceptance region using the hadron resonance gas
(HRG) model in  the canonical ensemble.
The HRG model is extended with the S-matrix corrections to the yields of protons and hyperons. 
The S-matrix calculation is based on the empirical phase shifts of
$\pi N$ scattering, an estimate of the $\pi \pi N$ background
constrained by Lattice QCD results of baryon-charge susceptibility,
and an existing coupled-channel model describing the $\vert S \vert
=1$ strange baryons.

It is demonstrated that an accurate description of the widths of resonances and the non-resonant interactions in a thermal model leads to a reduction of the proton yield relative to the HRG baseline (by $\approx 25\%$). Including the protons from strong decays of $\vert S \vert =1$ hyperons, which constitute $\approx 6\%$ of the total yields, does not alter this conclusion. 
Such a reduction is also crucial for resolving the proton anomaly in the LHC data.
For the hyperons, the S-matrix scheme predicts an increase in the $\Lambda+\Sigma^0$ yields relative to the HRG baseline by $\approx 23\%$.
This is consistent with the data from the ALICE collaboration.
Furthermore, the S-matrix prediction on the ratio of yields of proton to $\Lambda+\Sigma^0$ 
is in good agreement with the measured values by the  ALICE collaboration in Pb-Pb collisions in the events with the largest multiplicities ($dN_{ch}/d\eta$). The evolution of (multi-)strange baryons to $\Lambda$ yields with $dN_{ch}/d\eta$ calculated in the present thermal model in the C-ensemble 
follows the measured values within two standard deviations.

A good description  was obtained for the variation of the strangeness content in
the final state as a function of the
number of charged hadrons at mid-rapidity with  the same freezeout
temperature $T_f\sim 156.5$ MeV. This further supports the idea that
at LHC energies (independent of colliding system), the freezeout temperature  coincides  with the chiral crossover  as calculated in  LQCD. 
An exact conservation of strangeness is to be imposed in the full phase-space rather than in the experimental acceptance at mid-rapidity. Consequently,  the correlation volume parameter where strangeness is exactly conserved was found to be larger than the fireball volume  at mid-rapidity.

The S-matrix formulation discussed here
 accurately describes the measured hadron yields and supports
the interpretation of the LQCD results. It will be useful to employ this in analyzing the 
multi-strange baryon sectors and the light mesons.


\section{Acknowledgments}
K.R. acknowledges partial support  by the Polish National Science Center (NCN) under OPUS Grant No. 2018/31/B/ST2/01663,  and by the Polish Ministry of Science.  
N.S. acknowledges the support of SERB Research Scientist research grant (D.O. No. SB/SRS/2020-21/48/PS) of the Department of Science and Technology, Government of India. We also acknowledge  fruitful  collaboration with Pok Man Lo.

%
%
%

\bibliography{main}

\end{document}